\documentclass[a4paper,10pt,twoside]{cpc-hepnp}

\usepackage{multicol}
\usepackage{graphicx}
\usepackage{booktabs}
\usepackage{amssymb,bm,mathrsfs,bbm,amscd}
\usepackage[tbtags]{amsmath}
\usepackage{lastpage}

\newcommand{\KNsing}{$\{\bar KN\}_{I=0}$}
\newcommand{\KK}[1]{$\{K\bar K\}_{I=#1}$}

\begin{document}

\fancyhead[co]{\footnotesize D. Jido et al: A new $N^{*}$ resonance as a hadronic molecular state}

\footnotetext[0]{Received 14 March 2009}

\title{A new $N^{*}$ resonance as a hadronic molecular state
}

\author{%
      Daisuke Jido$^{1;1)}$\email{jido@yukawa.kyoto-u.ac.jp}%
\quad Yoshiko Kanada-En'yo$^{1}$%
}
\maketitle

\address{%
1~(Yukawa Institute for Theoretical Physics, Kyoto University,
Kyoto 606-8502, Japan)
}

\begin{abstract}
We report our recent work on a hadronic molecule state 
of the $K \bar K N$ system with $I=1/2$ and $J^{P}=1/2^{+}$. We 
assume that the $\Lambda(1405)$ resonance and the scalar 
mesons, $f_{0}(980)$, $a_{0}(980)$, are reproduced as 
quasi-bound states of $\bar KN$ and $K \bar K$, respectively. 
Performing non-relativistic three-body calculations
with a variational method for this system, we find a quasibound state
of the $K \bar K N$ system around 1910 MeV below the three-body 
breakup threshold. 
This state corresponds to a new baryon resonance of  $N^{*}$ 
with $J^{P}=1/2^{+}$. 
We find also that this resonance has the cluster structure of the two-body 
bound states keeping their properties as in the isolated two-particle systems.
We also briefly discuss another hadronic molecular state composed
by two $\bar K$ and one $N$, which corresponds to a $\Xi^{*}$ resonance. 
\end{abstract}



\begin{multicols}{2}

\section{Introduction}
\label{intro}

The structure of baryon resonances has been investigated in quark models,
in which symmetries of constituent quarks, such as spin, flavor and color,
and their radial excitations  
play a major role to describe the wavefunctions of the baryon resonances. 
Since the baryon resonances decay into low-lying mesons and a baryon 
with the strong interactions, the resonances may have also large components 
of mesons and baryons. For these components, inter-hadron dynamics 
is important and gives essential contributions to understand the structure 
of the baryon resonances.  

\begin{center}
\includegraphics[width=6.2cm]{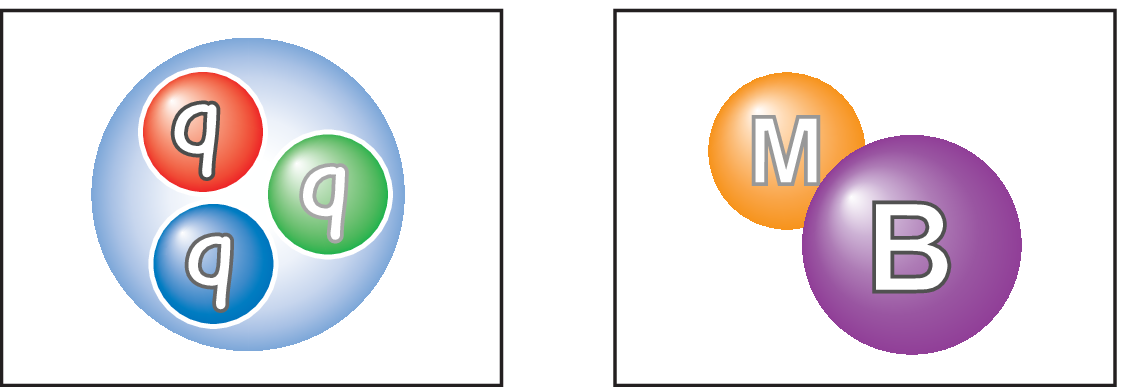}
\figcaption{Schematic pictures of baryon resonances.\label{fig1}} 
\end{center}

These two pictures are complementary, and which picture is realized in 
a baryon resonance depends on the energy scale to see the resonance,
since the interaction ranges are different in these pictures. Nevertheless, 
for some baryon resonances, the quark core components are not significant 
and the resonances are predominantly composed by hadrons. In such 
resonances, the quarks are clustered and their dynamics are 
confined inside the constituent hadrons. 

One of the historical examples is the $\Lambda(1405)$ resonance 
having $S=-1$ and $J^{p}=(1/2)^{-}$, which has been considered as a 
quasibound state of the $\bar KN$ system~\cite{Dalitz:1959dn}.
Modern calculations based on chiral dynamics with 
unitary coupled-channels formulations   
also reproduce the $\Lambda(1405)$ as a dynamically generated 
resonance in meson-baryon scattering with $S=-1$ and $I=0$~\cite{Kaiser:1995eg,Oset:1998it,Oller:2000fj,Oset:2001cn,GarciaRecio:2002td,Hyodo:2002pk,Jido:2003cb}.
A recent investigation~\cite{Hyodo:2008xr} pointed out that 
the $\Lambda(1405)$ can be regarded almost purely as 
a dynamically generated state in meson-baryon scattering,
while the description of the $N(1535)$ demands some components 
other than meson-baryon ones, such as genuine quark components. 
It has been also suggested that 
the $f_{0}(980)$ and $a_{0}(980)$ scalar mesons are molecular states
of $K \bar K$~\cite{Weinstein:1982gc}. 
The strong attraction in the $\bar K N$ system led to the idea
of deeply bound kaonic states in light nuclei, such as $K^-pp$ and 
$K^-ppn$, pointed out in Ref.~\citep{akaishi02}.
Later, many theoretical studies on the structure of the $K^-pp$ system 
have been performed, having turned out that the $K^-pp$
system is bound with a large width~\cite{Kpp}.
Two mesons and one baryon systems have been also investigated~\cite{MartinezTorres:2007sr,KanadaEn'yo:2008wm,Khemchandani:2008rk,Jido:2008kp,MartinezTorres:2008kh}, and one of 
their findings is that a quasibound state is formed in the  $K \bar K N$ 
system around 1910 MeV as an $N^{*}$ resonance~\cite{Jido:2008kp,MartinezTorres:2008kh}. 

In such multi-hadron systems, anti-kaon plays a unique
role for hadron dynamics due to its heavier mass and 
Nambu-Goldstone boson nature. 
According to the chiral effective theory, the $\bar KN$ and $\bar KK$ 
interactions are strongly attractive in the $s$-wave channel, 
and, owing to the heavy kaon mass, the $s$-wave interactions 
are more effective than those for the pion around the threshold energy.
In addition, being aware that the typical kaon kinetic energy in 
the bound systems estimated by the hadronic interaction range
is small in comparison with the kaon mass,
one may treat the kaons in multi-hadron systems in non-relativistic 
potential models done for nucleons in nuclear physics. 

Recent achievement of the studies on the $s$-wave $\bar K N$ 
effective potential is that $\bar{K}N$ interaction with $I=0$ is 
strongly attractive and develops a quasibound state somewhere 
around  the $\Lambda(1405)$ resonance. 
Starting from these strong $\bar KN$ interactions together with 
the $\bar K K$ quasibound picture of $f_{0}(980)$ and $a_{0}(980)$, we 
examine possible bound states of the lightest two-kaon nuclear systems
$K \bar K N$ and $\bar{K}\bar{K}N$ with $I=1/2$ and $J^P=1/2^+$, 
in the hadronic molecule picture. Using a non-relativistic potential 
model to describe the two-body interactions, we solve the three-body 
problem in a variational method, and find a new $N^{*}$ resonance 
which is not listed in Particle Data Table. 
The details of the calculation 
can be found in Refs.~\citep{KanadaEn'yo:2008wm,Jido:2008kp}.
Recently the $K\bar K N$ system was studied also in a more sophisticated 
calculation using the Faddeev formulation and they found a generated 
resonance at the same energy~\cite{MartinezTorres:2008kh, Albert} as our investigation.

\section{Formulation} \label{sec:formulation}

We use a non-relativistic three-body potential model for 
the $K\bar{K}N$ system.
The Hamiltonian for the $K\bar{K}N$ system is given by
\begin{eqnarray}\label{eq:hamiltonian}
H&=&T+V \\
V&\equiv& V_{\bar{K}N}(r_1)+V_{KN}(r_2)+V_{K\bar{K}}(r_3), 
\end{eqnarray} 
with the kinetic energy $T$ and 
the potential energy $V$ which consists of effective two-body interactions
given in $\ell$-independent local potentials as functions of 
$\bar{K}$-$N$, $K$-$N$ and $K$-$\bar K$ distances.
We assume isospin symmetry in the effective interactions, and we also
use isospin-averaged masses, $M_K=495.7$ MeV and $M_N=938.9$ MeV. 
We do not consider three-body forces nor transitions to two-hadron decays,
which will be important if the constituent hadrons are localized in a small region.

The effective interactions are described in complex-valued functions
representing the open channels, 
($\pi \Lambda$, $\pi \Sigma$) for $\bar K N$ and
($\pi \pi$, $\pi\eta$) for $K\bar K$. 
In solving Schr\"odinger equation for the three-body $K\bar{K}N$ system,
we first take only the real part of the potentials 
and obtain the wavefunctions in a variational approach
with a Gaussian expansion method developed in Ref.~\citep{Hiyama03}.
With the wavefunctions,  we calculate the bound state energies 
$E$ as expectation values of the total Hamiltonian~(\ref{eq:hamiltonian}).
The widths of the bound states are evaluated by the imaginary part of 
the complex energies as $\Gamma=-2\, {\rm Im}E$.

The effective interactions are parametrized in a one-ragne Gaussian 
form as
\begin{equation}
  V_{a}(r)  = U_{a} \exp\left[ - (r/b)^{2} \right] \label{eq:pot}
\end{equation}
with the potential strength $U$ and the range parameter $b$. 
The parameters used in this work are summarized in Table~\ref{tab:interactions}.

%\begin{table}[t]
\begin{center}
\tabcaption{\protect\label{tab:interactions} 
Parameters of the effective interactions and properties of two-body systems.
The energies ($E$) are evaluated from the corresponding two-body
breakup threshold. We also list the root-mean-square distances of the $\bar K N(I=0)$, $K\bar K(I=0)$ and $K\bar K(I=1)$ states,
which correspond to $\Lambda(1405)$ and $f_0(980),a_0(980)$, respectively.
For the $K\bar K$ interactions, we show the scattering lengths obtained 
in the present parameter. 
}
%\begin{center}
\begin{tabular}{ccc}
\hline\hline
 &\multicolumn{2}{c}{parameter set of interactions}\\
	&	(A)	&  (B)	\\
$b$ (fm)	&0.47 & 0.66 	\\
\hline
% & &  \\
$\bar K N$ & HW-HNJH & AY \\
\hline
$U^{I=0}_{KN}$ (MeV)	&	$-908-181i$	&	$-595-83i$	\\
$U^{I=1}_{KN}$ (MeV)	&	$-415-170i$	&	$-175-105i$	\\
% & & \\
$\bar K N(I=0)$ state	&		&		\\
Re$E$ (MeV)	&	-11 	&	-31 	\\
Im$E$ (MeV)	&	-22 		&	-20 	\\
$\bar K$-$N$ distance  (fm) 	&	1.9 	&	1.4 	\\
% & & \\
\hline
$K\bar K$	& KK(A)	&	KK(B)	\\
\hline
% & &  \\
%$b$ (fm)	&	0.47 		&	0.66 	\\
$U^{I=0,1}_{K\bar K}$ (MeV)	&	$-1155-283i$		&	$-630-210i$	\\
% & & \\
$K\bar K (I=0,1)$ state	&	&		\\
Re$E$ (MeV) 	&	-11 	&	-11 	\\
Im$E$ (MeV) 	&	-30 	&	-30 	\\
 $K$-$\bar K$ distance (fm) 	& 	2.1 		&	2.2 	\\
% & &  \\
\hline
$KN$	& KN(A)		&	KN(B)	\\
\hline
% & & \\
%$b$ (fm)	&   0.47 	&	0.66 	\\
$U^{I=0}_{KN}$ (MeV)	& 0 		&	0 	\\
$U^{I=1}_{KN}$ (MeV)	& 820 	&	231 	\\
% & & \\
$a^{I=0}_{KN}$ (fm)	& 	0 		&	0 	\\
$a^{I=1}_{KN}$	(fm) &$-$0.31 	&	$-$0.31 \\
% & & \\
\hline\hline
\end{tabular}
\end{center}
%\end{table}

One of the key issues for study of the $\bar KN$ interaction is the
subthreshold property of the $\bar KN$ scattering amplitude,
namely the resonance position of $\Lambda(1405)$.
Particle Data Group~\cite{PDG} reports 
the mass of the $\Lambda(1405)$ resonance around 1405 MeV, 
which is extracted mainly in the $\pi \Sigma$ final state interaction. 
Based on this fact, a phenomenological effective
$\bar K N$ potential (AY potential) was derived in Refs.~\citep{akaishi02,yamazaki07},
having relatively strong attraction in the $I=0$ channel to provide the $K^-p$ bound state at 1405 MeV. 
Recent theoretical studies of $\Lambda(1405)$ in coupled channels approach
with chiral dynamics have indicated that 
$\Lambda(1405)$ is described as a superposition of two pole states
and one of the states is considered to be a $\bar{K}N$ quasibound state 
embedded in strongly interacting $\pi\Sigma$ continuum~\cite{Jido:2003cb,hyodo07,Hyodo:2008xr}.  
This double-pole nature suggests  that the resonance position in the
$\bar{K}N$ scattering amplitudes with $I=0$ is around 1420 MeV, 
which is higher than the energy position of the 
nominal $\Lambda(1405)$ resonance. This is confirmed by
the bubble chamber experiments of the 
$K^{-} d \rightarrow n \Lambda(1405)$ reaction~\cite{Braun:1977wd}, 
in which the $\Lambda(1405)$ spectrum has a peak structure clearly 
at 1420~MeV instead of 1405~MeV. This reaction has been recently 
investigated in Ref.~\citep{Jido:2009jf}, and it was found that
the $\Lambda(1405)$ is produced by the $\bar KN$ channel 
in this reaction. 
Based on the chiral SU(3) coupled-channel dynamics,
Hyodo and Weise have derived another effective $\bar{K}N$ 
potential (HW potential)~\cite{hyodo07}.
The HW potential provides a $\bar{K}N$ quasibound state at 
$\sim$ 1420 MeV instead of 1405 MeV,
and is not as strong as the AY potential.
Here we compare these two different effective potentials. 
For the HW potential, we use the parameter set referred 
as HNJH in Ref.~\citep{hyodo07}, which was obtained by
the chiral unitary model with the parameters of Ref.~\citep{Hyodo:2002pk}.
We refer to this potential as ``HW-HNJH potential''

For the interactions of the $\bar K K$ systems with $I=0$ and $I=1$, 
the strengths $U_{\bar KN}$ in Eq.~(\ref{eq:pot}) are determined 
so as to form quasibound states 
having the observed masses and widths of $f_{0}(980)$ and $a_{0}(980)$.
We take the mass 980 MeV and the width 60 MeV as the inputs to
determine the $K\bar K$ interactions in both the $I=0$ and $I=1$ channels. 
In this phenomenological single-channel interaction, 
the effect of the two-meson decays such 
as $\pi\pi$ and $\pi\eta$ decays is 
incorporated in the imaginary part of the effective
$K\bar K$ interaction.
In this model, the $K\bar K$ 
interaction is independent of the total isospin of $K\bar K$,
because it is adjusted to 
reproduce the $f_{0}$ and $a_{0}$ scalar mesons having 
the same mass and width.
The repulsive $KN$ interaction is fixed by experimentally obtained 
scattering lengths: $a^{I=0}_{KN}=-0.035$ fm and 
$a^{I=1}_{KN}=-0.310\pm 0.003$~fm~\cite{KNint}. 
For the range parameters of the $\bar KK$ and $KN$ potentials we use
the same values of the $\bar KN$ interaction, $b=0.47$ fm 
for the HW-HNJH potential and $b=0.66$ fm for the AY potential.
The depths of the attractive potentials shown in Table~\ref{tab:interactions}
are compatible with the kaon mass. Nevertheless, the kinetic energies of the kaon 
in the two-body bound systems are small enough for nonrelativistic treatments 
of multi-kaon systems.
The two-body interactions are schematically summarized in Table~\ref{tab:2int}.
Hereafter we refer to the quasibound $\bar KN$ and $K\bar K$ states as \KNsing and \KK{0,1}, respectively.

\begin{center}
\tabcaption{Two-body interaction. The quasibound states 
in the corresponding channels are indicated as the resonances.
\label{tab:2int}}
\begin{tabular}{cccc}
\hline\hline
 & $I=0$ & $I=1$ & threshold \\
 \hline
 $\bar KN$ & $\Lambda(1405)$ & week attraction & 1434.6 MeV \\
 $K \bar K$ & $f_{0}(980) $ & $a_{0}(980)$ & 991.4 MeV \\
 $KN$ & very weak & strong repulsion & 1434.6 MeV \\
 \hline \hline
\end{tabular}
\end{center}

%%%%%

\section{Results} \label{sec:results}

Let us show the results of 
the three-body calculation of the $K \bar K N$ system with $I=1/2$ and 
$J^P=1/2^+$.  As shown in Fig.~\ref{fig:spe}, 
we find that, in both parameters (A) for HW-HNJH and 
(B) for AY, a $K\bar{K}N$ bound 
state is obtained below all of the threshold energies of 
the \KNsing$+K$, \KK{0}$+N$ and \KK{1}$+N$ channels, 
which correspond to the $\Lambda(1405)+K$, $f_0(980)+N$ and 
$a_0(980)+N$ states, respectively.
This means that the obtained bound state is stable against breaking up to
the subsystems.

\begin{center}
\includegraphics[width=6.5cm]{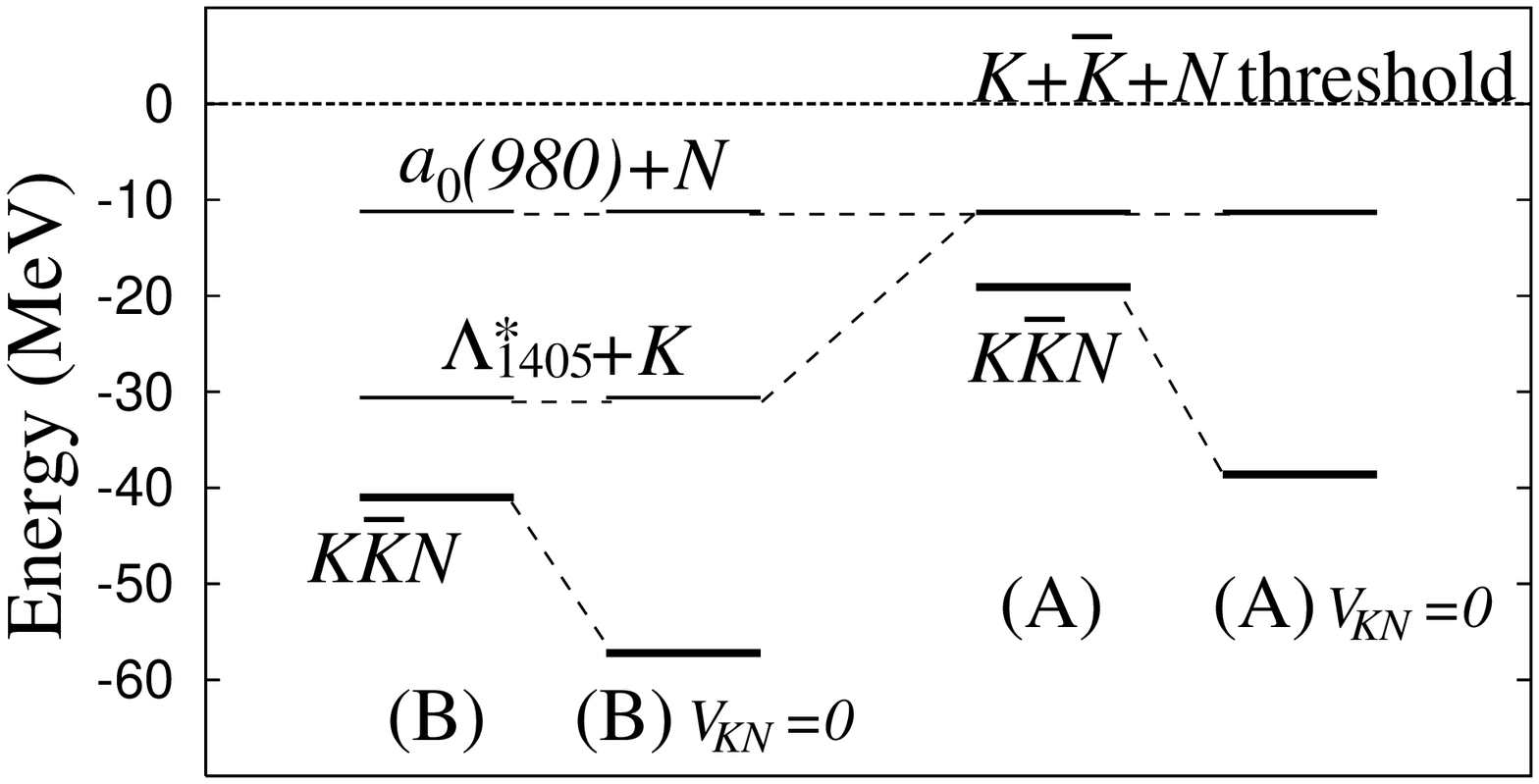}
\figcaption{\label{fig:spe} Level structure of the $K \bar{K}N$ system
calculated with (A)~the HW-HNJH potential
and (B)~the AY potential. The energies are measured from 
the $K$+$\bar K$+$N$ threshold located at 1930 MeV.
The bound state is denoted by $K \bar K N$.
The calculated  thresholds of the two-body decays to 
\KK{0,1}$+N$ and \KNsing$+K$ are denoted by
$a_{0}(980)+N$ and $\Lambda^{*}_{1405}+K$, respectively. 
The results obtained without the $KN$ repulsion are also shown. }
\end{center}

\subsection{Energy and width of the $K\bar KN$ state}

The values of the real and imaginary parts of the obtained energies are
given in~Table \ref{tab:energy}. The imaginary part of the energy 
corresponds to the half width of the quasi-bound state. 
The contribution of each decay mode is also shown 
as an expectation value of the imaginary potential  
$\langle {\rm Im} V \rangle$. 
The binding energies of the  $K \bar K N$ state measured 
from the three-body $K+\bar K+N$ threshold are
found to be $-19$ MeV and $-41$ MeV in the cases of 
(A) and (B), respectively. 
The difference stems from the fact that
the AY potential gives a deeper binding of the \KNsing\ state
than the HW-HNJH potential due to the stronger $\bar KN$ attraction. 
It is more physically important that
the $K \bar K N$ bound state appears about 10 MeV below
the lowest two-body threshold, \KNsing$+K$, 
in both cases (A) and (B). 
This energy is compatible to nuclear many-body system, 
and it is considered to be weak binding energy
in the energy scale of hadronic system.

%\begin{table}[t]
\begin{center}
\tabcaption{Energies of the $K\bar KN$ states 
calculated with parameter (A) and (B) given in Table~\ref{tab:interactions}. 
Contributions of $V^{I=0,1}_{\bar KN}$ and
$V^{I=0,1}_{K\bar K}$ to the imaginary energy are separately listed.
\protect\label{tab:energy} }
%\begin{center}
\begin{tabular}{lrr}
\hline\hline
parameter set	&	(A)	&	(B)	\\
$V_{\bar KN}$	&	HW-HNJH	&	AY	\\
\hline
Re$E$ 	&$	-19 	$&$	-41 	$\\
\ \ $\langle T \rangle$ &	169 	&	175 	\\
\ \ $\langle {\rm Re}V \rangle$ &	$-$188 	&	$-$216 	\\
\hline
Im$E$ 	&$	-44 	$&$	-49 	$\\
\ \ $\langle {\rm Im}V^{I=0}_{\bar KN} \rangle$ &$	-17 	$&$	-19 	$\\
\ \ $\langle {\rm Im}V^{I=1}_{\bar KN} \rangle$ &$	-1 	$&$	0 	$\\
\ \ $\langle {\rm Im}V^{I=0}_{K\bar K} \rangle$ &$	-1 	$&$	-4 	$\\
\ \ $\langle {\rm Im}V^{I=1}_{K\bar K} \rangle$ &$	-25 	$&$	-25 	$\\
\hline\hline
\end{tabular}
\end{center}
%\end{table}

The weakly bound system has the following significant feature.
Comparing the results of the $K \bar K N$ with 
the properties of the two-body subsystems shown in 
Table~\ref{tab:interactions}, it is found that
the obtained binding energies and widths of the $K \bar K N$ state
are almost given  by the sum of those of $\Lambda(1405)$ and 
$a_0(980)$ (or $f_0(980)$), respectively. 
This indicates that two subsystems, $\bar K N$ and $K\bar K$,
are as loosely bound in the three-body system as they are in two-body system.

The decay properties of the $K \bar K N$ state can be discussed by
the components of the imaginary energy.
As shown in Table~\ref{tab:energy}, 
among the total width $\Gamma=-2E^{\rm Im}\sim 90$ MeV, 
the components of the $\bar KN$ with $I=0$ and the $K\bar K$ with 
$I=1$ give large contributions
as about $40$ MeV and $50$ MeV, respectively.
The former corresponds to the $\Lambda(1405)$ decay to 
the $\pi\Sigma$ mode with $I=0$, while 
the latter is given by the $a_0(980)$ decay, which is dominated 
by $K\bar K \rightarrow \pi \eta$. 
In contrast, the $\bar KN$ $(I=1)$ 
and the $K\bar K$ $(I=0)$ interactions provide only small
contributions to the imaginary energy. This is because, as 
we will see later, the $\bar K N$ subsystem is dominated by the
$I=0$ component due to the strong $\bar KN$ attraction and 
the $K \bar K$ subsystem largely consists of  the $I=1$ component
as a result of the three-body dynamics. 
The small contributions of the $\bar KN$ $(I=1)$ 
and the $K\bar K$ $(I=0)$ interactions to the imaginary energy
implies that the decays to $\pi\Lambda K$ and $\pi\pi N$ are 
suppressed. 
Therefore, we conclude that the
dominant decay modes of the $K \bar K N$ state are 
$\pi\Sigma K$ and $\pi\eta N$.   This is one of the important 
characters of the $K \bar K N$ bound system. 

\subsection{Structure of the $K \bar K N$ state}

For the isospin configuration of the 
$K\bar{K}N$ state, we find that the $\bar K N$ subsystem has a dominant 
$I=0$ component, as shown in Table~\ref{tab:radii}.
In the $K\bar K$ subsystem, the $I=1$ configuration is dominant 
while the $I=0$ component gives minor contribution.
This is because, in both $I=0$ and $I=1$ channels, 
the $K\bar K$ attraction is equally strong enough to 
provide quasi-bound $K\bar K$ 
states, but the $I=1$ configuration of $\bar K K$ is favorable 
to have total isospin 1/2 for the $K \bar K N$ with the \KNsing\ subsystem. 
Due to this isospin configuration, the $K \bar K N$ system 
has the significant decay patterns as discussed above.

\begin{center}
\tabcaption{Isospin and spatial structure of the $K\bar KN$ state with 
parameter (A) and (B) given in Table~\ref{tab:interactions}. 
The r.m.s.\ radius of the $K$, $\bar K$ and $N$ distribution, and
the r.m.s.\ values for the $\bar K$-$N$, $K$-$\bar K$ and 
$K$-$N$ distances are listed. 
The detailed definitions are described in 
Ref.~\citep{Jido:2008kp}.
\protect\label{tab:radii} }
\begin{tabular}{lcc}
\hline\hline
	&	(A)	&	(B)	\\
	&	HW-HNJH	&	AY	\\
\hline
 \multicolumn{3}{c}{isospin configuration} \\
$\Pi \left(\left[\bar K N\right]_{0}\right)$	&	0.93 	&	0.99 	\\
$\Pi \left(\left[\bar K N\right]_{1}\right)$	&	0.07 	&	0.01 	\\
$\Pi \left(\left[K \bar K\right]_{0}\right)$	&	0.09 	&	0.17 	\\
$\Pi \left(\left[K \bar K\right]_{1}\right)$	&	0.91 	&	0.83 	\\
\hline
 \multicolumn{3}{c}{spatial structure}  \\
$r_{K\bar KN}$ (fm) & 1.7   & 1.4   \\
$d_{\bar KN}$ (fm)	&	2.1 		&	1.3 	\\
$d_{K\bar K}$ (fm)	&	2.3 	 &	2.1 	\\
$d_{KN}$ (fm)	&	2.8 	 &	2.3 	\\
\hline\hline
\end{tabular}
\end{center}

We discuss the spatial structure of the $K \bar KN$ bound system. 
In Table \ref{tab:radii}, we show the root-mean-square (r.m.s.) 
radius of $K\bar K N$, $r_{K \bar K N}$, and 
r.m.s.~values for the $\bar K$-$N$, $K$-$\bar K$ and $K$-$N$
distances, $d_{\bar KN}$, $d_{KN}$, $d_{K\bar K}$.
The definitions are given in Ref.~\citep{Jido:2008kp}.
The r.m.s.~distances  of the two-body systems, 
\KNsing\ and  \KK{0,1}, are shown 
in Table~\ref{tab:interactions}.
It is interesting that the present result shows that 
the r.m.s.~$\bar K$-$N$ and $K$-$\bar K$ distances in the three-body
$K\bar KN$ state have values close to those in the quasi-bound two-body 
states, \KNsing\ and  \KK{0,1}, respectively.
This implies again that the two subsystems of the three-body state
have very similar characters with those in the isolated two-particle systems.

Combining the discussions of the isospin and spatial structure of the
$K\bar KN$ system, we conclude that the structure of the $K \bar KN$
state can be understood  
simultaneous coexistence of $\Lambda(1405)$ and $a_{0}(980)$
clusters as shown in Fig.~\ref{fig:bond}. This does not mean
that the $K \bar KN$ system is described as superposition of 
the $\Lambda(1405)+K$ and $a_{0}(980)+N$ states, because
these states are not orthogonal to each other. The probabilities for the 
$K \bar K N$ system to have these states are 90\%. It means that $\bar K$ is shared by 
both $\Lambda(1405)$ and $a_{0}$ at the same time.

\begin{center}
\includegraphics[width=3.2cm]{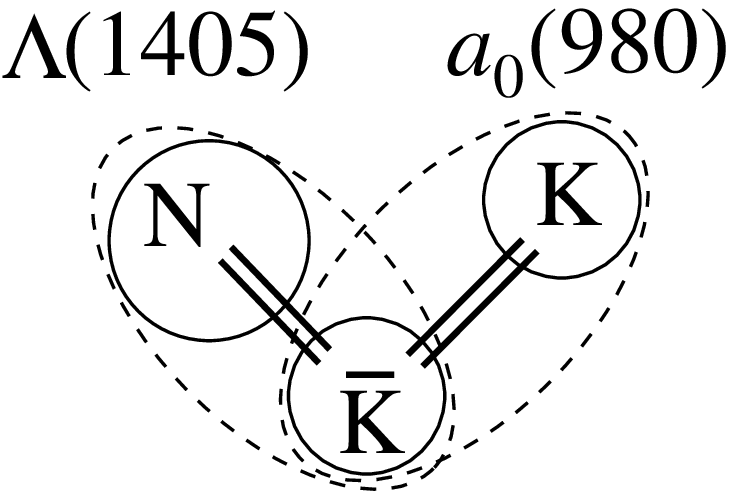}
\figcaption{Schematic structure of the $K\bar KN$ bound system.\label{fig:bond}} 
\end{center}

It is also interesting to compare the obtained $K \bar KN$ state with 
nuclear systems. 
As shown in Table~\ref{tab:energy}, the hadron-hadron distances in the 
$K \bar KN$ state are about 2 fm, which is as large as typical
nucleon-nucleon distance in nuclei. 
In particular, in the case~(A), the hadron-hadron distances are larger than 2 fm and 
the r.m.s.~radius of the three-body system is also  as large as 1.7 fm. This is
larger  than the r.m.s.\ radius 1.4 fm of $^4$He.
If we assume uniform sphere density of the three-hadron system
with the r.m.s.~radius 1.7 fm, the mean hadron density
is to be 0.07 hadrons$/$(fm$^{3}$). Thus the $K\bar KN$ state
has large spatial extent and dilute hadron density.

We discuss the role of the $KN$ repulsion in the
$K \bar K N$ system. In  Ref.~\citep{Jido:2008kp}, we have shown the results 
calculated without the $KN$ interaction. There it has been found
 in both  (A) and (B) cases 
that  the binding energy of the $K \bar KN$ state is 20 MeV larger than 
the case  with the $KN$ repulsion,
and that the widths also becomes larger, $\Gamma=130-140$ MeV. 
We also obtain spatially smaller three-body system. 
As a result of the localization, 
the system can gain more potential energy and larger imaginary energy
in the case of no $KN$ interaction than the case with the $KN$ repulsion. 
In other words, thanks to the $KN$ repulsion, 
the $K \bar K N$ state is weakly bound  and its width is suppressed to be
as small as the sum of the widths of the subsystems. 
The distances of the two-body subsystems obtained without the $KN$
interaction are as small as about 1.5 fm, which is comparable with the sum of 
the charge radii of proton (0.8 fm) and $K^{+}$ (0.6~fm).
For such a small system, 
three-body interactions and transitions to two particles could be 
important. In addition,  
the present picture that the system is described in nonrelativistic three
particles might be broken down, and one would need relativistic 
treatments and  two-body potentials with consideration of 
internal structures of the constituent hadrons.

Finally we shortly discuss the $\bar K \bar K N$ system with $S=-2$,
$I=1/2$ and $J^{p}=(1/2)^{+}$.
For the $\bar K \bar K N$ system, the binding energy from 
the $\Lambda(1405)+\bar K$ threshold is found to be as small as a few MeV
due to the strong repulsion $\bar K \bar K$ with 
$I=1$~\cite{KanadaEn'yo:2008wm}. The reason 
of the small binding energy is understood by isospin configuration of this
system. Due to the strong attraction of $\bar KN$ with $I=0$, 
one of the $\bar K N$ pair forms a quesibound $\Lambda(1405)$ state.
At the same time, the other pair of $\bar KN$ has dominantly 
$I=1$ component to have total isospin $1/2$ of the $\bar K \bar K N$ 
system. Although the $\bar KN$ with $I=1$ is attractive, 
the attraction is not enough to overcome the repulsive $\bar K \bar K$
interaction.

\section{Conclusion and discussion} \label{sec:summary}

We have investigated the $K \bar KN$ system with $J^{p}=1/2^{+}$ 
and $I=1/2$ in the non-relativistic three-body calculation under 
the assumption that $\bar KN$ and $K \bar K$ systems form quasibound
states as $\Lambda(1405)$, $f_{0}(980)$ and  $a_{0}(980)$.
The present three-body calculation suggests 
a weakly quasibound state  below all threshold
of the two-body subsyetems. 
For the structure of the $K\bar KN$ system,
we have found that  
the subsystems of  $\bar KN$ and $K\bar K$ 
are dominated by  $I=0$ and $I=1$, respectively,
and that these subsystems have very similar 
properties with those in the isolated two-particle systems. 
This leads that the $K \bar KN$ quasi-bound system
can be interpreted as coexistence state of $\Lambda(1405)$
and $a_{0}(980)$ clusters, and $\bar K$ is a constituent
of both $\Lambda(1405)$
and $a_{0}(980)$ at the same time. 
Consequently, the binding energy and the width of 
the $K\bar KN$ state 
is almost the sum of those in $\Lambda(1405)$ and $a_{0}(980)$,
and  the dominant decay modes are
$\pi \Sigma K$ from the $\Lambda(1405)$ decay 
and $\pi \eta N$ from the $a_{0}(980)$ decay.
The decays to $\pi \Lambda K$ and $\pi\pi N$ channels are 
suppressed. 
We also have found that the root-mean-square radius of the  
$K\bar KN$ state is as larger as 1.7 fm and the inter-hadron 
distances are larger than 2 fm. These values are comparable to,
or even larger than, the radius of $^{4}$He and typical 
nucleon-nucleon distances in nuclei. Therefore, 
the $K\bar KN$ system more spatially extends 
than typical baryon resonances.  
These features are caused by the weak binding of the three hadrons,
for which the $KN$ repulsive interaction plays an important role. 

Our finding that the $\Lambda(1405)$ keeps its properties 
in few-body systems motivates the $\Lambda(1405)$ doorway 
picture for the $\bar K$ absorption into nucleus discussed in 
Ref.~\citep{Sekihara:2009yk}, in which the non-mesonic decay
of kaonic nuclei is investigated under the assumption that 
the $\bar K$ absorption takes place through the $\Lambda(1405)$.

For the experimental confirmation of the new $N^{*}$ resonance discussed 
here, a recent paper~\cite{MartinezTorres:2009cw,Oset} discusses 
a relation of the new $N^{*}$ state with the bump structure in 
$\gamma p \to K^{+}\Lambda$ observed in recent 
experiments~\cite{Bradford:2005pt}, and proposes other experimental
consequences in different reactions.
 
Although the obtained $K \bar K N$ state is located below
the thresholds of  $\Lambda(1405)+K$, 
$f_0(980)+N$ and $a_0(980)+N$, 
there could be a chance to access the $K \bar K N$ state energetically  
by observing the $\Lambda(1405)+K$, $f_0(980)+N$ and $a_0(980)+N$ 
channels in the final states,  because these resonances have as large widths as  
the  $K\bar K N$ state. 
Since, as we have shown, the $K \bar K N$ state
has the large $\Lambda(1405)+K$ component, 
the $K \bar K N$ state could be confirmed in its decay to  $\Lambda(1405)+K$
by taking coincidence of the $\Lambda(1405)$
out of the invariant mass of $\pi\Sigma$ and the three-body 
invariant mass of the $\pi\Sigma K$ decay.
It is also interesting point that 
the $K \bar KN$ state can be a doorway state of
the $\Lambda(1405)$ production. 
This means that  presence of 
the three-body $N^{*}$ resonance at 1.9 GeV could affect 
spectrum of $\Lambda(1405)$ production.
Thus, this $N^{*}$ would explain the 
strong energy dependence of the $\Lambda(1405)$ production 
observed in the $\gamma p \to K^{+} \pi^{\pm}\Sigma^{\mp}$ reaction 
at SPring8~\cite{Niiyama:2008rt}.

\acknowledgments{
The authors would like to thank Professor~Akaishi,
Dr.~Hyodo, Dr.~Dot\'e, Professor~Nakano and Professor~J.K.~Ahn
for valuable discussions. 
This work is supported in part by
the Grant for Scientific Research (No.~18540263 and
No.~20028004) from Japan Society for the Promotion of Science (JSPS)
and from the Ministry of Education, Culture,
Sports, Science and Technology (MEXT) of Japan.
A part of this work is done under Yukawa International Project for 
Quark-Hadron Sciences (YIPQS).
The computational calculations of the present work were done by
using the supercomputer at YITP.}

\vspace{-2mm}
\centerline{\rule{80mm}{0.1pt}}

\end{multicols}
\clearpage

\end{document}